\documentclass[12 pt]{article}
\usepackage[utf8]{inputenc}
\usepackage{authblk}
\usepackage{epigraph}
\usepackage{hyperref}
\usepackage{graphicx}
\usepackage{mathtools}   
\usepackage{amsfonts}
\usepackage{appendix}
\usepackage{caption}
\usepackage{orcidlink}
\graphicspath{ {./figures/} }
\hypersetup{
    colorlinks=true,
    linkcolor=blue,
    urlcolor=blue,
}

\usepackage[style=apa,url=false,sorting=nyt,natbib=true]{biblatex}
\providecommand{\keywords}[1]
{
  \small	
  \textbf{\textit{Keywords---}} #1
}

\title{{GPS Observables in Newtonian Spacetime or Why We Do Not Need `Physical' Coordinate Systems}\thanks{Cite as: Mozota Frauca, \'A. GPS Observables in Newtonian Spacetime or Why We Do Not Need ‘Physical’ Coordinate Systems. \textit{Euro Jnl Phil Sci} 14, 51 (2024). \url{https://doi.org/10.1007/s13194-024-00611-7}}}

\author[1]{Álvaro Mozota Frauca}
\affil[ ]{alvaro.mozota@udg,edu, \orcidlink{0000-0002-7715-0563} \href{https://orcid.org/0000-0002-7715-0563}{https://orcid.org/
0000-0002-7715-0563}}
\affil[1]{Department of Physics. Universitat de Girona. C/ de la Universitat de Girona, 1. 17003 Girona (Spain)}

\date{\today}

\begin{document}

\maketitle

\begin{abstract}
Some authors have defended the claim that one needs to be able to define `physical coordinate systems' and `observables' in order to make sense of general relativity. Moreover, in \citep{Rovelli2002a}, Rovelli proposes a way of implementing these ideas by making use of a system of satellites that allows defining a set of `physical coordinates', the GPS coordinates. In this article I oppose these views in four ways. First, I defend an alternative way of understanding general relativity which implies that we have a perfectly fine interpretation of the models of the theory even in the absence of `physical coordinate systems'. Second, I analyze and challenge the motivations behind the `observable' view. Third, I analyze Rovelli's proposal and I conclude that it does not allow extracting any physical information from our models that wasn't available before. Fourth, I draw an analogy between general relativistic spacetimes and Newtonian spacetimes, which allows me to argue that as `physical observables' are not needed in Newtonian spacetime, then neither are they in general relativity. In this sense, I conclude that the `observable' view of general relativity is unmotivated.
\end{abstract}


\section{Introduction}

In \citep{Rovelli2002a} Carlo Rovelli proposes an ingenious construction that would allow the construction of a ‘physical coordinate system’ and therefore a definition of ‘physical observables’ in general relativity. Physical observables are thought to be needed in order to have a complete interpretation of general relativistic models and in order to extract all their gauge invariant content. In this article I will challenge this view and argue that we have a perfectly fine interpretation of general relativistic models and that we are able to extract all of their physical content.

The core of my disagreement with Rovelli and other authors like Rickles and Earman lies in whether the diffeomorphism invariance or general covariance\footnote{Later on I will comment on the different ways these terms are used and distinguished.} of general relativity is just a formal feature or whether it has some physical meaning and deeper implications. This debate is an old one, starting right after the formulation of general relativity, when Kretschman objected \citep{kretschmann_uber_1917} that since any physical theory can be formulated in a generally covariant manner, then general covariance must be void of any physical meaning. Authors like Rovelli reject Kretschmann's objection and argue that general covariance has deep implications that make it the case that the nature of physical magnitudes in general relativity is fundamentally different from the nature of physical magnitudes in other theories such as Newtonian mechanics. I will argue against this view and argue that the nature of physical magnitudes is the same in both theories.

The position of these authors about the nature of physical magnitudes is motivated by the difficulty of expressing the physical content of general relativity in a way that is independent of representational choices such as coordinate systems and in a way that identifies the same physical content in diffeomorphism-related models\footnote{An historical predecessor of these views is \citep{Bergmann1961}.}. These authors defend that one needs to define `observables', which are thought to be these invariant quantities that capture the physical content of general relativity. In order to do so, one needs to define `physical coordinate systems', which are thought to be internal reference systems that allow describing the evolution of other variables not in terms of arbitrary coordinates or points on a manifold, but with respect to these internal degrees of freedom.

%
In this article I will take the `physical observable' view to consist of the following three claims\footnote{The discussion below will show how the different authors endorse these claims to different degrees. In particular, while Rovelli \citep{Rovelli1991,Rovelli2002a,Rovelli2004} clearly argues for the two first claims, he is more ambiguous about the third. Meanwhile Rickles and Earman \citep{Earman2002,Earman2006,rickles_chapter_2008} clearly argue for claim 3.}:
\begin{enumerate}
\item{The construction of `physical observables' and `physical coordinate systems' is necessary for having a complete interpretation of general relativity and for being able to extract the physical content of a general relativistic model.}
\item{General covariance forces us to abandon the idea of external or idealized reference systems. Therefore, it forces us to explicitly introduce internal reference systems.}
\item{General covariance implies that there are deep ontological differences between general relativity and other spacetime theories.}
\end{enumerate} 
The main goal of this article is to argue against these three claims. I start in section \ref{sect_intro_general_relativity} by offering a positive account of my view. That is, by outlining the ways in which I think that one can interpret general relativity and extract all its physical content in a way that does not need to explicitly define `physical observables'. Then, in section \ref{sect_need_for_obs} I expand on the `physical observable' view and the motivations behind it.  In the rest of the article I further argue against the three claims in two ways. First, in section \ref{sect_GPS_observ_GR} I analyze the GPS construction proposed by Rovelli and I argue that all the physical content it allows us to extract was already available in the original model that didn't include the `physical observables' he defines. Second, in section \ref{sect_Newtonian_differential_geometry} I establish an analogy with Newtonian physics which weakens the case for observables, and then in section \ref{sect_replies} I analyze the possible replies of the defender of observables to my arguments. Other authors have argued against some of these claims in some different ways \citep{Maudlin2002,Pons2010,pooley_background_2017,Pitts2017,Pitts2018,read_classical_2023} and my arguments can be seen as complementary to theirs.

\section{General relativity and its interpretation}\label{sect_intro_general_relativity}


The diffeomorphism invariance or general covariance of general relativity is a formal feature of the theory that has motivated important conceptual discussions about its physical content and how to define it. While authors like Rovelli have argued that we need to find `observables', i.e., diffeomorphism invariant quantities, in order to extract all the physical information of a general-relativistic model and have a complete understanding of it, in this section I will argue that this is not the case. That is, I will argue that we have a perfectly fine grasp of the physical content of general-relativistic models which consists of a causal structure, a geometry, and an inertial structure. We are able to read this from our models with no need to introduce `observables' or `physical coordinate systems.'


\subsection{A brief discussion of how to interpret general relativity}\label{sect_gr_interpretation}

Let me start this section by giving a definition of a general relativistic model. It is given by a 4-dimensional manifold $\mathcal{M}$, a Lorentzian metric $g_{\mu\nu}$ defined on it, and by (possibly) some matter and force fields, here represented by $\phi$. In some more technical definitions of spacetime models\footnote{See for instance the definitions in \citep{Kuchar1980,Malament2012,Knox2014} of Newtonian models or the recent comparison in \citep{Meskhidze2023} of general relativistic models and models of teleparallel gravity, which essentially differ in the connections employed in each of these models.} one also includes the affine connection $\nabla$ as one of the ingredients of the model, but here I will just assume that the connection for general relativistic models is the Levi-Civita connection and I won't be paying much more attention to it. Note also that my discussion is compatible with other choices for the connection. In this sense, a model will be given by the triple $\langle \mathcal{M}, g_{\mu\nu},\phi \rangle$. For it to be a valid general relativistic model, it has to satisfy the Einstein equations that relate the curvature of the metric tensor with the stress-energy tensor of the fields, which satisfy their own equations of motion.

The basic physical interpretation of a general relativistic model has three basic tenets. First, the metric tensor $g_{\mu\nu}$ defines the causal structure of spacetime. At every point the metric tensor distinguishes three types of vectors: spacelike, timelike, and null. Causal curves in spacetime are the ones that have timelike or null tangent vectors, and material bodies are only allowed to follow such curves. Similarly, the equations of motion of matter and force fields are built in a way that respects the causal structure of spacetime.

The second aspect of the physical interpretation of the model is the chronogeometric or chronometric meaning of the metric tensor and its relation with what clocks measure. The metric tensor defines a geometry for spacetime, and, in particular, it defines a proper time $T$ along any timelike curve $X^{\mu}(\tau)$:
\begin{equation}\label{proper_time}
T=\int d\tau \sqrt{-g_{\mu\nu}\dot{X}^{\mu}\dot{X}^{\mu}} \, ,
\end{equation} 
where I am adopting the sign convention $\{-,+,+,+\}$ for the metric and $\dot{X}^{\mu}$ represents the derivative of the coordinate $X^{\mu}$ with respect to $\tau$, the (arbitrary) parameter parametrizing the curve. This proper time is interpreted as the time that a physical system moving along this trajectory would experience. When the system is a clock, this implies that the reading of the clock will be proportional to its proper time. This is known as the clock hypothesis, and it has the status of an assumption or of a postulate because at this level of discussion we are not giving an account of how the dynamics of any physical system or of any clock picks up the proper time. In this sense, it is an assumption about how physical systems behave in spacetime.

At this point it is relevant for our discussion to emphasize that general relativity is a theory that defines possible geometries of spacetime and that it is not a theory about how we get to know about this geometry, that is, it is not a theory about how our clocks work. In this sense, it makes predictions about what clocks would measure without needing to explicitly include them in our models. That is, the model does not care about the details of our clocks to make the right predictions for them, as long as they remain close to ideal clocks, i.e., clocks which always `measure' their proper time\footnote{Point-particles could act as ideal clocks, and indeed experiments with muons have been used for testing the hypothesis. Extended material objects can only act as ideal clocks in as far as we can approximate the physical processes happening in them as point-like and occurring along a time-like trajectory. \citep{fletcher_light_2013} shows how it is possible in principle to build a light clock that approximates ideal clocks to a desired accuracy.}. To insist on this point, let me compare a general relativistic model with a world map. By reading a map and knowing the scale (that is different at different points on the map) we can learn about the geometry of our planet, e.g., about the distance between two points or about the shortest path between London and Paris. The map is just a description of the geometry and doesn't tell us how to measure distances and in this sense it also comes with an analog of the clock hypothesis: we are assuming that our distance measurements, no matter how we perform them, will agree with the distances predicted by the map. General relativistic models, just as maps, define geometries but do not tell us how to measure them.

There are different views on how to understand the clock hypothesis and the relation between the geometry of spacetime and the dynamics of physical systems, including clocks. Some authors \citep{Maudlin2012} take geometry to be fundamental and to explain the behavior of physical systems and clocks. Some other authors \citep{Brown2006} take an opposing view: it is the dynamics of physical systems and fields that explains why the field $g_{\mu\nu}$ can be interpreted in a geometric way. For the discussion in this article the distinction between the geometric and dynamical approaches to general relativity won't be very relevant, as both seem to agree on the interpretation of the models in that $g_{\mu\nu}$ defines a geometry and a notion of proper time and in that physical systems and clocks dynamically correlate with this proper time. These two shared pieces of interpretation will be enough to establish the main claims in this article.

A consequence of this is that I intend to remain neutral in this article with respect to the debate between the geometric and dynamical views of general relativity. Even if the presentation in this article is in geometric terms, I believe that the vast majority of my claims will remain true if expressed or interpreted in the dynamic view.

Before moving on to the next point, let me note that in discussions of spacetime theories besides discussing the role of clocks in measuring time, it is common to discuss the role of rods in measuring distances. Following other authors\footnote{See for instance \citep{Maudlin2012}.}, I won't be introducing any `rod hypothesis' or paying much attention to spatial distances, as they are not naturally defined in general relativity and they are not necessary for our analysis.

Finally, the third aspect of the interpretation of general relativity has to do with the way it defines inertial motion, i.e., the trajectories that free bodies would follow, which in the case of general relativity are timelike geodesics for bodies with non-zero mass and null geodesics for bodies with zero mass, such as photons. This is just a generalization of Newton's second law to curved spacetimes. That is, if we want to predict how a body will move in a general relativistic spacetime, we do just as in Newtonian mechanics: we study to which forces it is subject and this tells us how much it will deviate from inertial motion. There are two main differences between the inertial structure of general relativity and Newtonian mechanics. One is that in Newtonian mechanics a body moving under gravitational influence is a body subject to force and hence deviates from inertial motion, while in general relativity gravity is not a force, and its effect is encoded in the inertial structure of spacetime. That is, in general relativity free-falling bodies follow inertial trajectories and are not subject to any force. The second difference with the Newtonian case is that the inertial structure of spacetime is dynamical, that is, different spacetimes will have different inertial structures and in different regions of spacetime inertial behavior may be different. 

\subsection{Diffeomorphism invariance}

Having briefly introduced general relativistic models and the way we interpret them for extracting what they say about the world we are in a position to discuss diffeomorphism invariance. If we have a general relativistic model $\langle \mathcal{M}, g_{\mu\nu},\phi \rangle$ satisfying Einstein field equations, we can build an equivalent model by applying a diffeomorphism, i.e., an invertible, differentiable map from $\mathcal{M}$ to itself. This maps any point $P$ to another one $P'$, and one can transform appropriately\footnote{This transformation is known as a pull-back.} $g_{\mu\nu}$ and the matter fields $\phi$ so that the properties at $P'$ are the same as they originally were at $P$. For instance, if a timelike curve $\gamma$ on $\mathcal{M}$ is mapped to another one $\gamma'_{\mu\nu}$, the proper time along the new curve as computed using (\ref{proper_time}) but with the transformed metric $g'$ is the same as the proper time along the original curve computed using the original metric $g_{\mu\nu}$. In this sense, the two models $\langle \mathcal{M},g_{\mu\nu},\phi \rangle$ and $\langle \mathcal{M},g'_{\mu\nu},\phi' \rangle$ describe the same physics but just changing the points in the manifold that represent a given physical event. We can again compare with the case of world maps, the two models $\langle \mathcal{M},g_{\mu\nu},\phi \rangle$ and $\langle \mathcal{M},g'_{\mu\nu},\phi' \rangle$ are just like two different maps, possibly using different projections for describing the geometry of the Earth.

At this point let me mention that a diffeomorphism is conceptually different from a change of coordinates, but that this distinction is subtle and can be ignored in many discussions of general relativity. The reason for this is that the distinction between a point and the coordinates we use for referring to it won't be very relevant and in practice it won't make a difference to claim that an event $E$ originally represented by point $P$ with coordinates $x^{\mu}$ is now represented by the same point but with different coordinates $y^{\mu}$ or that it is now represented by a point $P'$ that has coordinates $y^{\mu}$. For this reason, one may use the terms diffeomorphism and change of coordinates indistinctively, and loosely say that two diffeomorphism-related models are related by a change of coordinates in the context of general relativity\footnote{We will later see that in other contexts one can define diffeomorphism transformations slightly differently so that with this definition diffeomorphisms are not equivalent to changes of coordinates. For a discussion of all these subtleties I refer the reader to \citep{pooley_background_2017}.}.  

Diffeomorphisms define equivalence classes of models $\langle \mathcal{M},g_{\mu\nu},\phi \rangle$ in which every two models are related by a diffeomorphism. In this sense, every triple $\langle \mathcal{M},g_{\mu\nu},\phi \rangle$ in the equivalence class represents the same physical situation. For instance, if a triple $\langle \mathcal{M},g_{\mu\nu},\phi \rangle$  contains a timelike curve connecting a point $P_0$ with a point $P_f$ in a proper time $T$ which could represent the trajectory of the Earth around the Sun during one year, in any other triple $\langle \mathcal{M},g'_{\mu\nu},\phi' \rangle$ in the equivalence class there will also be a timelike geodesic connecting a pair of points $P'_0$ and $P'_f$ in the same proper time $T$ and with the same physical properties as described by $\phi'$ along the trajectory. To insist on the analogy, the equivalence class of models $\langle \mathcal{M},g_{\mu\nu},\phi \rangle$ is analogous to the collection of all the maps that represent the geometry of our planet. In the same way that reading any world map one can learn about the distance between Paris and London, taking any triple in the equivalence class one can learn about the geometry of a possible general relativistic spacetime.

Diffeomorphism invariance and the fact that we have to deal with equivalence classes of models have given rise to a number of conceptual debates, and in particular to the debate I am focusing on this paper. The position I am defending here is that the diffeomorphism invariance of general relativity does not mean that the physical content of this theory is fundamentally different from the physical content of other spacetime theories, such as Newtonian spacetime. Indeed, I will argue that the way Newtonian spacetime models and general relativistic models are to be interpreted in analogous ways, which are the ways I have outlined in this section and which I will come back to in section \ref{sect_Newtonian_differential_geometry}. In this sense, I will disagree with authors like Rovelli who claim that we need to introduce `physical' coordinate systems in order to understand general relativity. That is, I will argue that general relativistic models get a clear physical interpretation by means of what has been discussed in this section and that their physical content is perfectly well-defined even in the absence of `physical' coordinate systems\footnote{An alternative way of putting my claim is that any coordinate system in general relativity is equally physical.}. In the next section I will introduce and analyze the arguments for positions contrary to mine.

Before this, let me clarify that I will take this debate to be mostly independent of other debates in the foundations of general relativity. As I have already mentioned, there is a debate between two views of how to understand the relationship between dynamics and geometry which I take to be somewhat orthogonal to this debate. That is, I believe that when I claim that the physical content of a general relativistic model is clear, this holds independently of whether one takes geometry to be more fundamental than dynamics or not. Similarly, the authors I will be opposing do not necessarily take a stance in these debates.

The debate between spacetime substantivalism and relationalism will be more relevant for the discussion in this article. The reason for this is that some of the arguments for the `physical observables' view have a clear relationalist motivation. However, I want to make clear that one can reject that view without committing to substantivalism. That is, I will argue that one can claim that the physical content of general relativity is clear with no need to introduce `physical observables' both from substantivalist and relationalist positions.

The diffeomorphism invariance of general relativity has been used as an argument against spacetime substantivalism.  The fact that two diffeomorphism-related models describe different states of affairs for a point $P$ in the manifold, or for a coordinate point $x^{\mu}$, has been argued to signal a failure of determinism or as an argument against spacetime substantivalism. This is the famous hole argument that was first formulated by Einstein in 1913 and which didn't receive much attention until it was put forward again in \citep{Earman1987}. However, the consensus in the philosophy of physics community seems to be that the diffeomorphism invariance of general relativity still leaves room for a substantivalist position known as sophisticated substantivalism, which is able to combine substantivalist intuitions with the fact that diffeomorphism-related models are equivalent in a way that does not imply indeterminism\footnote{See \citep{Hoefer1996,pooley_hole_2006,Norton2019}.}.

We will see how the arguments by Rovelli are connected to the hole argument, as he mentions \citep[Sect. 2.2, 2.3]{Rovelli2004} it as a motivation for his own view. However, while I will argue against Rovelli's view, the discussion in this article won't reanalyze this particular argument in detail. In this sense, I will be aligning with the consensus in the community and claim that the diffeomorphism invariance of general relativity supports either a relationalist or a sophisticated substantivalist view, but I won't agree with Rovelli in that it further implies the need for introducing `physical observables'. 


\section{The need for physical observables?}\label{sect_need_for_obs}

In the introduction, I described the ‘physical observables view’ as consisting of three claims:
\begin{enumerate}
\item{The construction of `physical observables' and `physical coordinate systems' is necessary for having a complete interpretation of general relativity and for being able to extract the physical content of a general relativistic model.}
\item{General covariance forces us to abandon the idea of external or idealized reference systems. Therefore, it forces us to explicitly introduce internal reference systems.}
\item{General covariance implies that there are deep ontological differences between general relativity and other spacetime theories.}
\end{enumerate} 
In this section, I will explain the two main motivations behind the view.

\subsection{Gauge and `observables'} \label{sect_gauge_observables}

The analysis of general relativity using the structures of gauge theory introduced the notion of `observable' in the debates on the foundations of general relativity. However, this term is used with different meanings. In the first meaning of the term, observables are the `physical quantities that we can predict and measure in real experiments' \citep[p. 1]{Rovelli2002a}. Then, there is a technical sense of observable which is any quantity, in the canonical formulation of general relativity, that has vanishing Poisson brackets with the constraints of the theory. There is yet another technical meaning, which is any quantity that can be used for characterizing equivalence classes under diffeomorphisms of general relativistic models.

For general gauge theories like electromagnetism, there is a correspondence between the first and second meaning: physical quantities like the electromagnetic field are represented by phase-space functions that have vanishing Poisson brackets with the constraints\footnote{Some authors \citep{Pitts2017} have questioned this technical definition of observable and formulated an alternative one in which observables are required to have vanishing Poisson brackets with a function known as the generator. Taking this alternative view wouldn't affect the discussion in this article.}. The argument then is that in order to determine the physical content of a theory (its observables in the intuitive sense), one needs to find these phase space functions (its observables in the technical sense). As in the case of general relativity there is no general construction that allows defining the observables in this technical sense, Rovelli, Rickles, and Earman conclude that we have a problem also with observables in the intuitive sense\footnote{See the following: \citep[p. 1]{Rovelli2002a}, \citep[pp. 299,315]{Rovelli1991}, \citep[Sect. 2.4.3]{Rovelli2004}, \citep{Earman2002,Earman2006,rickles_chapter_2008}.}. In this way they derive claim 1. Add to this that they believe that explicitly adding reference systems can solve the problem and you have claim 2. Finally, when you consider that general relativity can be formulated as a gauge theory and other spacetime theories cannot (in principle), then you can conclude something similar to claim 3.

From my point of view, this line of argument is mistaken, as I believe that we do have a good understanding of general relativity even if we do not have well-defined observables in the technical sense. As I have discussed in the previous section, we are able to give a sensible interpretation of equivalence classes of models $\langle \mathcal{M},g_{\mu\nu},\phi \rangle$ and we are able to extract all of their physical content. To insist, this content includes the geometry of spacetime, which can be used to derive the behavior of clocks and bodies moving in spacetime. Contrary to what Rovelli claims, we are able to extract all of the physical content of our models, and we are not `far from capturing all the physics' \citep{Rovelli2002a}. In this sense, one is applying a formal recipe beyond its domain of applicability to arrive at a mistaken conclusion.

Several authors have argued against this gauge analysis of general relativity\footnote{See \citep{Maudlin2002} for a clear conceptual discussion and \citep{Gryb2010,Thebault2012,Gryb2016,Pons2010,Pitts2017,Pitts2018,MozotaFrauca2023} for some complementary discussions.}, given that the structure of gauge transformations like the ones in electromagnetism and the diffeomorphism invariance of general relativity are very different. In particular, gauge transformations in electromagnetism can be thought of locally: we can apply a transformation at a spacetime point $x^{\mu}$ which transforms the 4-potential but which leaves the electromagnetic field at that point unchanged. In this sense, it makes sense to claim that the value of the field at the point $x^{\mu}$ remains invariant under such a local transformation. In the case of general relativity, such a local view is not available. Under a diffeomorphism the fields at a coordinate point $x^{\mu}$ become the pullback of the fields that were originally at some other coordinate point $x'^{\mu}$. Now it doesn't really make sense to claim that what is invariant at the point $x^{\mu}$ is the value of the physical fields at that point as one could do in the electromagnetic case. The reason for this is that the coordinate point $x^{\mu}$ represents two different spacetime points before and after the diffeomorphism and it does not make sense to compare two different points in order to look for what is invariant or the physical content of the theory at a spacetime point. 

Moreover, for some simple toy models it is easy to see\footnote{See the discussion of the double harmonic oscillator model in \citep{MozotaFrauca2023}.} that no non-trivial `observables' in this technical sense can be defined, while one has a perfectly fine interpretation of these models. One could expect the same conclusion to be true for the case of general relativity. For these reasons, I agree with the authors cited that such `observables' aren't needed for understanding general relativity and that it is likely that they are not even well-defined. As I have already stated, the goal of this article is not to analyze the technical details of the phase-space structure of general relativity and its relation with gauge transformations but to argue that we do have a clear interpretation of this theory and that we are able to extract all of its physical content.

Finally, notice that if one accepts this sort of argument, one is also led to accept that there is a distinction between general relativity and other spacetime theories such as Newtonian mechanics. As mentioned above, this distinction can be taken to imply just that we need to explicitly introduce reference objects as part of our general relativistic models, or something stronger such as that the nature of spacetime models is fundamentally different in the case of general relativity. I will come back to the issue of the gauge interpretation of general relativity in section \ref{sect_replies}.

\subsection{Relationalism and `observables'}\label{sect_relationalism}

The second motivation that Rovelli and other authors have for introducing `physical observables' or `physical coordinate systems' is a strong version of relationalism and a deep skepticism with respect to spacetime talk. More precisely, Rovelli rejects that coordinates in a manifold or in spacetime have physical meaning unless there is a physical system that measures them. In this sense, Rovelli believes that it does not make sense to describe the trajectory of a body in terms of coordinates in a manifold and that one should describe it in relation to some physical reference fields or objects. Similarly, one cannot speak about the value of a field at a coordinate point, but should instead speak about the value of the field when some other fields take given values. Rovelli concludes that the physical content of general relativity is contained in correlations between physical magnitudes which would be the observables we can empirically observe.

It is beyond the scope of this paper to precisely characterize Rovelli's version of relationalism\footnote{I refer the interested reader to Rovelli's discussion in \citep[Chap. 2, 3]{Rovelli2004}  and to Thebault's characterization of this sort of relationalism in \citep{Thebault2012,Thebault2021}.}. For our discussion here it will be enough to notice that it is of a radical sort that requires that reference objects are explicitly included in our modeling, which implies that if there isn't any reference object we wouldn't be able to extract physical predictions from our models. The concept of reference object is understood widely here, as any field could act as a reference object.

It is straightforward to see how this sort of radical relationalism fits well with claims 1 and 2 above. The construction of observables in the technical sense of the term is for Rovelli the way to define the correlations between physical fields and objects in which they believe the physical content of general relativity lies. For this reason, in order to extract all these correlations, and hence the physical content of the model, they argue that one needs to define these physical observables. As this content lies in relations including reference objects, one needs to explicitly introduce them in our models.

As one can infer from my discussion in section \ref{sect_intro_general_relativity}, I reject this sort of relationalism and its consequences, or, at least, I argue that there are ways of understanding general relativity that do not need to commit to this view and that they allow us to have a sensible interpretation of the models and to extract all their physical content. In the first place, relationalist views do not need to be as radical as Rovelli's relationalism. Even if one agrees that spacetime models encode the relations between different fields or bodies, or in between them and some reference objects, this does not mean that one needs to explicitly introduce these objects in our models to be able to extract predictions about them. 

Consider one of the examples discussed by Rovelli \citep[p. 72]{Rovelli2004}. In this example one has a clock on the surface of the Earth and another one orbiting in a circular orbit around the planet. These two clocks would be interchanging light signals with their readings and therefore they would be recording the correlations $T_1(T_2)$ and $T_2(T_1)$, that is, the time measured by clock 1 as received by clock 2 when its proper time was $T_2$ and the other way around. These correlations are part of the physical content of the model as anyone in this debate agrees. Now, one can be a relationalist and think that these correlations are part of this physical content while some other parts of the model (those regarding spacetime geometry for instance) are meaningless unless more reference objects are present. Even in this case, one can reject claims 1 and 2 for this example. That is, one can extract the predictions $T_1(T_2)$ and $T_2(T_1)$ without defining `observables' and without explicitly introducing the clocks in the model. One just needs to specify the two time-like trajectories and the model will determine the proper times the clocks will measure, the null trajectories followed by the signals, and finally the correlations $T_1(T_2)$ and $T_2(T_1)$. At no moment is it necessary to add explicitly the dynamics of the clocks, as the clock hypothesis and the assumption that light signals travel along null geodesics suffice for extracting the predictions of the model.

For this reason, even if one believes that a spacetime model is a way of encoding relations like $T_1(T_2)$ and $T_2(T_1)$ between actual systems in the world, one can still claim that one does not need to explicitly include these physical systems in our models in order to extract predictions about them. 

Relationalists can also take a view about spacetime models in which they are taken to encode not only relations between actual physical systems but between possible ones that do not need to be actually there. That is, predictions like $T_1(T_2)$ and $T_2(T_1)$ count as part of the physical content of the model also in a counterfactual way: they represent the recordings that the two clocks would register if they were there in spacetime. All these counterfactual predictions that can be extracted from a spacetime model are very intuitively part of their physical content. In this sense, someone with relationalist intuitions does not need to commit to the same kind of relationalism as Rovelli and accept claims 1 and 2.

In the case of substantivalist views, the situation is similar. Clearly, the substantivalist agrees that counterfactual predictions are part of the physical content of a general relativistic model. That is, predictions like the time a clock would register or the trajectory a free body would travel are part of the physical content of a general relativistic model for the substantivalist. Moreover, the substantivalist will claim that the physical content of the model includes geometrical properties of spacetime that the relationalist would only accept as encoding the relations (actual or potential) between material bodies. In this sense, the standard substantivalist position is to reject Rovelli's radical relationalism and claims 1 and 2.

In this sense, I believe one can reasonably reject Rovelli's radical relationalist perspective, and take some other well-established alternatives. For this reason, the radical relationalist motivation for the position I am opposing in this article can be rightly challenged.

Finally, I want to note that nothing in the radical relationalist motivation is specific about general relativity. Indeed, in some passages by Rovelli in which he discusses the interpretation of Newtonian mechanics, he seems to adopt a radical relationalist perspective, as I will further discuss in section \ref{sect_replies}. This creates some tension between the two motivations for his view, as the gauge perspective pulls in the direction of interpreting Newtonian mechanics and general relativity differently, while his relationalist view pulls in the opposite direction. I will exploit this tension in this article to argue against the three claims above and I will come back to this issue in section \ref{sect_replies}.

\section{GPS observables in general relativity} \label{sect_GPS_observ_GR}

In this section I want to move from the general discussion in the previous section to consider a concrete implementation of Rovelli's ideas. In his 2002 article \citep{Rovelli2002a}, Rovelli introduces GPS observables as an example of how to define observables (in both the technical and intuitive senses) which would allow us to extract the physical content of general relativistic models. In this section I will introduce this construction and I will argue that, against Rovelli's view, it is not needed to make a sensible interpretation of general relativistic models, as even without it we are able to make predictions following what I discussed in section \ref{sect_intro_general_relativity}.

\begin{figure}[ht]
\centering
\includegraphics[width=0.4\textwidth]{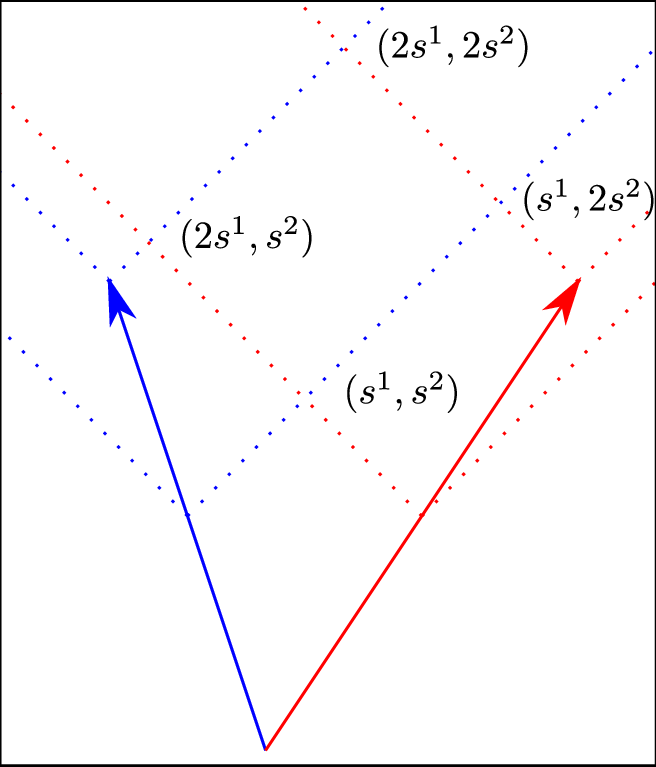}
\caption{\label{figure_gps_coord} Representation of how GPS coordinates are assigned to spacetime points for a 2-dimensional spacetime. Each satellite (in red and blue) carries its own clock and at each moment of time it sends a signal (dotted lines) into space with the reading of its clock. For a region of spacetime $\mathcal{R}$, the signals $s^{\alpha}$ can individuate spacetime points, i.e., to each set of coordinates, to each set of signals, there corresponds a unique spacetime point. This construction works both in Newtonian and relativistic spacetimes, even if the rules describing the evolution of clocks and light signals are different in different spacetimes.}
\end{figure}

In Rovelli's construction there are four satellites sending signals into space. He proposes that each of the four satellites would be starting at the same spacetime point and that they would carry a clock that would be measuring their proper time $s^{\alpha}$, where $\alpha=1,2,3,4$ is an index that identifies the satellite. Each satellite, at each moment in its trajectory would send a signal into space indicating the reading of its clock at the time it emits the signal. From each spacetime point in a region $\mathcal{R}$ one would receive four signals corresponding to the four satellites, and these can be used to identify each spacetime point in the region. In this sense, the signals sent by the satellites define a coordinate system and we can express all physical facts about the region $\mathcal{R}$ using this coordinate system. For instance, we can express the metric tensor in this coordinate system $g_{\mu\nu}(s^{\alpha})$ and the same holds for the value of any matter field $\phi(s^{\alpha})$. This idea is represented in figure \ref{figure_gps_coord} in a two-dimensional version.

The coordinate system $s^{\alpha}$ has the virtue that it is precisely defined and that agreeing to use it allows identifying spacetime points unambiguously. That is, it does not matter if one is using a model $\langle \mathcal{M}, g_{\mu\nu},\phi \rangle$ or a diffeomorphism-related one $\langle \mathcal{M}, g'_{\mu\nu},\phi' \rangle$, that the predictions, such as the value of a field $\varphi$, for a spacetime point identified by $s^{\alpha}$ will be the same. This is in contrast with the case in which one doesn't specify the way coordinates individuate spacetime points: if I use the model $\langle \mathcal{M}, g_{\mu\nu},\phi \rangle$ I could make the claim that at the spacetime point with coordinate $x^{\mu}$ the field $\varphi$ takes a value $\varphi(x^{\mu})=\varphi_1$, while if I use the model $\langle \mathcal{M}, g'_{\mu\nu},\phi' \rangle$, the coordinate $x^{\mu}$ will identify a different spacetime point\footnote{Assuming of course that we keep fixed the coordinate system. That is, if the point $P$ in the manifold gets a coordinate $x^{\mu}$ in the first model, in the second model the manifold point $P$ is still labeled with the same coordinate $x^{\mu}$.} and the value of the field $\varphi$ at that coordinate point will be a different one $\varphi(x^{\mu})=\varphi_2.$ In this sense one can claim that while $\varphi(s^{\alpha})$ is independent of the choice of representative in the equivalence class\footnote{In the literature one finds claims that $\varphi(s^{\alpha})$ is diffeomorphism invariant, but one has to be careful with the way this idea is phrased as it may refer to slightly different notions in different contexts and I prefer to stick to the way I am phrasing it here. In particular, in the Hamiltonian context `diffeomorphism invariance' may refer to functionals that are independent of the leaf and foliation in which they are evaluated, and this meaning is not the one I have in mind here.}, $\varphi(x^{\mu})$, with unspecified $x^{\mu}$, isn't. 

It is for this kind of reasoning that Rovelli claims that $s^{\alpha}$ are physical coordinates while some other $x^{\mu}$ wouldn't be, and that quantities like $\varphi(s^{\alpha})$ would be `observables' of the theory and $\varphi(x^{\mu})$ wouldn't. Indeed, the `observables' of general relativity, those quantities one would need to find to understand the theory, would just be the metric and matter fields expressed in this coordinate system. 

I agree with Rovelli in that by making use of this construction and coordinate system one is able to extract the physical content of a general relativistic model, at least for a region of spacetime. The disagreement is about whether it is necessary or not. The question we should ask ourselves is therefore whether we are gaining anything beyond a practical way of assigning coordinates to spacetime points at the time of interpreting the theory. My answer is that we are not, as, leaving aside the prediction that there will be signals traveling through the region, there is no prediction in the model that includes the satellites and signals that wasn't there in the model without the satellites and signals.

For instance, the model with the satellites and a field $\varphi$ predicts a distribution of values of the field $\varphi (s^{\alpha})$, while the model with no satellites predicts a diffeomorphism-related distribution $\varphi (x^{\mu})$. My claim is that these two predictions are perfectly equivalent in what concerns the field distribution. Imagine that the field distribution is such that in a region $\mathcal{R}$ the field takes $n$ extremal values, i.e., maxima and minima. Two diffeomorphism-related distributions will assign different coordinates to these $n$ points, but it will agree that there are $n$. Furthermore, if we connect these points using geodesics (for simplicity) the spacetime model determines the distance\footnote{Distance understood as space-like, time-like, or null depending on the type of geodesic. In certain spacetimes there may be more than one geodesic joining two points, but let me leave these technicalities aside for the sake of the discussion.} between them along this geodesics. In this sense, any diffeomorphism-related model describes the same distribution of field values and the same geometry. Similarly, these diffeomorphism-related models predict the trajectories that free bodies would follow. Even if in arbitrary coordinates, the prediction is completely equivalent to the prediction done using the coordinates $s^{\alpha}$. For instance, for a given trajectory the value of the field along the trajectory as a function of the proper time $\varphi (\tau)$  is the same for any diffeomorphism-related model. Similarly, if we have different trajectories that meet at different points the proper time elapsed in between each of these meeting points, according to any of the trajectories, is also obtained independently of the coordinate system used. 

We could go on and add light signals between trajectories, other fields, bodies moving under the influence of forces, and so on. The point is that the physical content of general relativistic models is perfectly clear in any arbitrary coordinate system and with no need to define `physical coordinate systems'. To insist, for any member of an equivalence class of models under diffeomorphisms, we are able to define a geometry of spacetime, predict what clocks would read along time-like trajectories, and deduce how free bodies would move. Adding operational ways of assigning coordinates to spacetime points doesn't add much to the picture, and does not allow us to deduce some physical content from our models that we could not deduce before assigning these coordinates.

There is a sense in which adding the satellites and signals adds new predictions to our model. Of course, if there is no satellite sending signals, the coordinate system $s^{\alpha}$ wouldn't correlate with what physical detectors in the region would detect. In this sense, the correlation $\varphi (s^{\alpha})$ is a new prediction of the model, as with no satellites and signals there is no correlation that we could measure. However, there is a strong sense in which $\varphi (s^{\alpha})$ was already a prediction of the original model. The reason for this is that the model allows computing the proper time that any body would experience along any time-like trajectory, and also the trajectory of any light signal. In this sense, we can use the coordinate system $s^{\alpha}$ even without explicitly adding the satellites and signals to our model, and indeed, even in the case in which these satellites aren't there in the actual world.

This connects with my discussion of radical relationalism in the previous section. Rovelli argues that the clocks, the satellites, and the signals need to be explicitly included in the model and he seems to imply that if these clocks weren't there or if we didn't introduce them in our model, the coordinates $s^{\alpha}$ would be meaningless. As I argued in the previous section, one can argue against this claim, as counterfactual reasoning allows us to connect them with the behavior of satellites and signals if they happened to be in the world. While Rovelli would reject the claim that spacetime coordinates could acquire meaning in this way, it seems to me that one can reasonably disagree with him on this point. And to insist, the interpretation of general relativistic models remains the same and complete, independently of the coordinate system used for describing it.

Let me clarify that of course there is a sense in which physical clocks and reference systems have to be included in a general relativistic setting: as material bodies they curve spacetime, and, strictly speaking, spacetime would be different if they weren't there. However, in many physical situations and to a very good degree of approximation one can reasonably neglect the effects that the gravitational influence of clocks, rods, or satellite signals may have. Granting that clocks and rods will have a gravitational effect, there is no reason not to treat them externally and ignore the effects they produce, just as we introduce other, and far cruder, approximations in our models.


Implicit in the claim that we need GPS coordinates or some other set of `physical' coordinates is that other coordinate systems are meaningless. However, that claim is also too strong. For simple and symmetric spacetimes it is clear that one can connect coordinate systems with what ideal clocks and rods would read. For instance, the standard time coordinate in Minkowski spacetime can be associated with the proper times of stationary clocks, and the spatial coordinates can be associated with the proper times that a series of light signals take to travel between certain clocks. Similarly, the standard $t,r$ coordinates in a Schwarzschild spacetime outside the horizon can also be connected with the proper time of stationary clocks and with the redshift in a family of light signals. For more generic spacetimes, the physical meaning of coordinates is not that straightforward, but it seems that the clock hypothesis of general relativity allows us to connect any coordinate system with the way in which we would expect that ideal clocks, possibly exchanging signals, would behave in such spacetimes. Notice also that the physical meaning of coordinates in any spacetime is given by the behavior that ideal clocks \textit{would} show, but that we do not need to have actual, physical clocks in spacetime in order to claim that these coordinates are meaningful. In this sense, all the coordinate systems are equally physical, as we are able to connect them with the geometry of spacetime and the behavior of reference objects.

Finally, notice that Rovelli provides an argument for showing that GPS observables are observables in the intuitive sense of the term, but not in the technical one. To prove that the GPS observables are observables in this sense would amount to building functions in the phase space of general relativity (plus satellites and signals) which correspond to $g_{\mu\nu}(s^{\alpha})$ and which have vanishing Poisson brackets with the constraints of general relativity. Indeed, given that the GPS coordinate system is generically just valid for some region of spacetime, it seems unlikely that one can build such phase space functions with all the desired properties.

In any case, even if we granted that the GPS observables could be constructed as invariant phase space functions, do we really want to claim that we need to construct them in order to have an understanding of general relativity? In particular, do we really need to explicitly add clocks (and satellites, and light signals) in our models to understand them? As far as I can see, we do not need GPS observables, phase space invariant functions, or `physical' coordinates in order to make sense of general relativistic models. The discussion of general relativity in the language of Hamiltonian mechanics can be technically involved and the comparison with gauge theory may be tempting, but at the end of the day I side with the philosophers of physics and physicists who have argued that we have a perfectly fine understanding of general relativity in terms of equivalence classes of models under diffeomorphisms. That is, we do not have any trouble in reading out of a general relativistic model the causal structure of spacetime, the way clocks would behave, and the way free bodies would move in such a spacetime. When we add matter, there is no trouble in reading predictions about the distribution of matter or about the configuration of fields from our models.

The discussion in this section should have made clear that constructions like the GPS observables do not show that there is some problem with the ways of understanding general relativity I have described in section \ref{sect_intro_general_relativity}, i.e., the ways of understanding general relativity which do not require of `physical coordinate systems' or of `physical observables'. Similarly, one can resist the claim that reference objects should be treated internally to the theory. In the next section I will further argue for these claims by introducing an analogy with Newtonian spacetime and by expressing it in the language of differential geometry. The analogy will help establish that claim 3 is wrong, in that it is very plausible to argue that spacetime models, general relativistic or not, are interpreted in analogous ways for every kind of spacetime theory. Furthermore, it will allow me to argue against claims 1 and 2, as I will argue that the demand to build `physical coordinate systems' does not really make sense in either case, given that in both cases we are able to understand our models by making use of coordinate systems, arbitrary or not, and some sort of clock hypothesis that links them with what our clocks measure.

\section{Newtonian spacetime in the language of general relativity}\label{sect_Newtonian_differential_geometry}

General relativity and its coordinates are usually compared with the case of Newtonian physics and its absolute space and time. In this sense, while in Newtonian physics time and space coordinates are physical and meaningful, in general relativity this wouldn't be the case and one would need constructions like the GPS coordinates expressed above to make sense of the physics encoded in the generally covariant models. In this section I will be arguing against this view by pointing to the fact that Newtonian physics can be expressed in the language of general relativity and that coordinates in this case also get physical meaning by means of a clock hypothesis and similar hypotheses about how light signals travel through spacetime or about how ideal rods behave. In this sense, our understanding and interpretation of Newtonian models is the same as in general relativity: coordinates are as physical and meaningful in the one case as in the other, as in both cases the way they acquire physical meaning is by some assumption linking them and the metric structures of the models with the behavior of ideal systems. 

Furthermore, I will note that to understand Newtonian physics we do not need to introduce ideal clocks (nor actual ones), and it is enough to know how they would behave. In this sense, even if in a Newtonian setting one could rehearse arguments that are similar to Rovelli's to argue for the construction of GPS observables or some similar set of coordinates, it is clearly not the way physicists think about Newtonian models, and they are able to make sense of the models without explicitly introducing any clocks and rods in them. The analogy therefore shows that if one does not need to explicitly introduce reference objects in our models in Newtonian physics, one shouldn't expect to need them in the case of general relativity.

My argument in this paper could have been based on some other spacetime, such as Galilean spacetime\footnote{See \citep{Malament2012,Knox2014} for differential-geometrical descriptions of Galilean spacetime.}, but I have chosen Newtonian spacetime for convenience, given that it will later allow me to easily define an analog of the GPS observables. In Newtonian spacetime there are absolute time and absolute space, and this can be modeled in the language of differential geometry by a triple $\langle \mathcal{M}, t_{\mu}, h_{\mu\nu} \rangle$. $\mathcal{M}$ is a 4-dimensional manifold, and the forms $t_{\mu}$ and $h_{\mu\nu}$ encode the spatiotemporal structure, in a slightly different way to the way $g_{\mu\nu}$ encodes spacetime geometry for relativistic spacetimes\footnote{As commented above, I won't be discussing the role that an affine connection could be playing in these geometrical models. I refer the reader to \citep{Malament2012} and references therein for an introduction to Newton-Cartan theory, a way of codifying Newtonian gravity in the affine connection of a spacetime model.}.

The form $t_{\mu}$ is associated with time\footnote{There are alternative differential-geometrical formulations of Newtonian spacetime that use a temporal metric tensor $t_{\mu \nu}$ instead of a one form $t_{\mu}$. For consistency, there are some conditions that $t_{\mu}$ needs to satisfy, such as being non-vanishing and continuous.} and it describes the temporal structure of spacetime. First, it determines whether a vector is spacelike or timelike. If a vector $v^{\mu}$ satisfies $t_{\mu}v^{\mu}\neq 0$ then it is timelike, otherwise it is spacelike. This is analogous to the causal structure of general relativistic models, as we are able to define a causal trajectory in spacetime as one that follows a timelike curve, i.e., one with an everywhere-timelike tangent vector. This causal structure is just the causal structure of Newtonian spacetime, as we are able to define simultaneity surfaces, i.e., to foliate spacetime into a sequence of spaces, defined to be surfaces (3-dimensional regions of $\mathcal{M}$) in which any two points can be connected by spacelike curves. Second, $t_{\mu}$ also defines a metric for time, i.e., it defines absolute time. To compute the time along a trajectory in spacetime $X^{\mu}(\tau)$ we find an expression analogous to \ref{proper_time}:
\begin{equation}\label{proper_time_Newtonian}
T=\int d\tau t_{\mu}\dot{X}^{\mu} \, .
\end{equation} 
By postulating that clocks measure this time, we have the analog of the clock hypothesis for Newtonian spacetime. However, contrary to the case of general relativity, one can show that this `proper' time is independent of the trajectory in spacetime chosen. That is, if instead of choosing a timelike trajectory $X^{\mu}$ in between initial and final spacetime points we chose another one $X'^{\mu}$ and computed the `proper' time along that trajectory using \ref{proper_time_Newtonian} we would obtain the same result. Moreover, this `proper' time is independent of where in a simultaneity surface a trajectory starts or ends, it only depends on the initial and final simultaneity surfaces. In this sense, we see how $t_{\mu}$ contains the information of absolute Newtonian time: it defines simultaneity surfaces and an absolute measure of time between them that is independent of any spacetime trajectory. Ideal clocks are objects that no matter their trajectory in spacetime, their readings are proportional to their proper time \ref{proper_time_Newtonian}, which is nothing but absolute time.

$h_{\mu\nu}$ completes the geometrical information about spacetime by providing a geometry for space. We are defining space to be absolute, and the degenerate metric $h_{\mu\nu}$ encodes the structure of absolute space. $h_{\mu\nu}$ defines a spatial geometry for space, which allows defining lengths, areas, angles, and so on and it also defines a notion of absolute rest.


$t_{\mu}$ and $h_{\mu\nu}$ together codify the geometry of Newtonian spacetime in any arbitrary coordinate system using the language of differential geometry. Together with this geometry and causal structure, we can define inertial motion in Newtonian spacetime by postulating that free bodies move in straight lines in space at uniform velocities\footnote{In particular, one can postulate that the trajectories of free bodies minimize the action $S=\int d\tau \frac{h_{\mu\nu}\dot{X}^{\mu}\dot{X}^{\nu}}{t_{\mu}\dot{X}^{\mu}}$. These trajectories are straight lines at uniform velocities for the most natural definitions of the connection.}. By adding matter and force fields $\phi$ we would have a complete model of Newtonian physics. Such a model has an invariance under diffeomorphisms similar to general relativity. That is, given a model $\langle \mathcal{M},t_{\mu},h_{\mu\nu},\phi \rangle$ we can build an equivalent one by means of a diffeomorphism that transforms $t_{\mu},h_{\mu\nu}$, and $\phi$\footnote{Some authors choose not to call this transformation `diffeomorphism', and prefer to use the term for transformations that affect only dynamical variables and not fixed variables. However, I prefer sticking to a terminology that can be applied to kinematical models independently of their dynamical interpretation. For a discussion of the different notions of diffeomorphism transformation and diffeomorphism invariance and their relation with the differences in the interpretation of general relativity and other models I refer the reader to \citep{pooley_background_2017}. }. Just as in general relativity, the physical content of the model is represented equally well by any member of the equivalence class of models under diffeomorphisms. 

The interpretation of Newtonian models is similar to the interpretation of general relativistic models. For the latter, in section \ref{sect_gr_interpretation} I have discussed how the metric tensor encodes three features of spacetime: causal structure, metric structure, and inertial structure. Similarly, in Newtonian models these three features are encoded by $t_{\mu}$ and $h_{\mu\nu}$, as I have just discussed. In this sense, we read from $\langle \mathcal{M},t_{\mu},h_{\mu\nu},\phi \rangle$ the same three pieces of geometric and physical information as we did from $\langle \mathcal{M},g_{\mu\nu},\phi \rangle$. In table \ref{Table_GR_NewtonianST} I compare both theories in a way that highlights the analogies.


\begin{table}
\begin{tabular}[t]{lcc}
\hline
 & General Relativity & Newtonian spacetime                              \\ \hline
Model                       & $\langle \mathcal{M}, g_{\mu\nu},\phi \rangle$                           & $\langle \mathcal{M},t_{\mu}, h_{\mu\nu},\phi \rangle$  \\ 
Causal structure             &  Spacelike $g_{\mu\nu}\dot{x}^{\mu}\dot{x}^{\nu}>0$                         & Spacelike/Simultaneous $t_{\mu}\dot{x}^{\mu}=0$ \\
             &  Timelike $g_{\mu\nu}\dot{x}^{\mu}\dot{x}^{\nu}<0$                         & Timelike $t_{\mu}\dot{x}^{\mu}\neq 0$ \\
             &  Null $g_{\mu\nu}\dot{x}^{\mu}\dot{x}^{\nu}=0$                         &  \\
Clocks            &         $\int d\tau (-g_{\mu\nu}\dot{X}^{\mu}\dot{X}^{\mu})^{1/2}$              &     $\int d\tau t_{\mu}\dot{X}^{\mu}$     (or $t$)                   \\
Rods             &                      &           $\int d\tau (h_{\mu\nu}\dot{X}^{\mu}\dot{X}^{\nu})^{1/2}$    (or $x,y,z$)             \\ 
Free body             &  Timelike geodesic                         & Straight line, uniform velocity                          \\
Light             &  Null geodesic                         & Straight line, velocity = $c$                          \\
Dynamical geometry?             &  Yes                         & No                          \\
\hline
\end{tabular}
\caption{Comparison between general relativistic and Newtonian spacetimes.}
\label{Table_GR_NewtonianST}
\end{table}

An important similarity in the interpretation of Newtonian models is that we also need a clock hypothesis to connect coordinates and geometric tensors ($t_{\mu}$) with what clocks measure. Similarly, we could formulate a rod hypothesis that connects the coordinates and the tensor $h_{\mu\nu}$ with what rods measure. Alternatively, if one introduces assumptions about how light signals behave one can make operational constructions similar to the ones employed in special relativity. That is, if one assumes that light moves at uniform velocity $c$ with respect to absolute space, one can just operationally define the distance between two static clocks to be half the time it takes for a light signal to travel from one clock to the other and back divided by $c$. In this way, the operational notions of time and space in Newtonian spacetime become clearly analogous to those in special relativity, as in both cases we would be relying on a clock hypothesis and some assumptions about how light signals behave.

Formulating the geometry of Newtonian spacetime in the language of general relativity helps highlight the similarities between our spacetime theories and puts some pressure on the idea that coordinates or `observables' are radically different in the case of general relativity. To this kind of argument one expects authors like Rovelli to reply\footnote{Take for instance Rovelli and Vidotto's claim that: `the theory [general relativity] is written in terms of spacetime coordinates $x$ and $t$, but the physical meaning of these is totally different from the physical meaning of the spacetime coordinates with the same name used in special relativity and in non-relativistic physics. The spacetime coordinates $X$ and $T$ in non-relativistic and special relativistic physics have metric meaning: the spacetime coordinates $x$ and $t$ in general relativistic physics do not have metric meaning' \citep[p. 7]{Rovelli2022}.} that the difference in the case of Newtonian spacetime is that in this spacetime there is a privileged set of coordinates, $t,x,y,z$, which would correspond to absolute time and absolute distance along three orthogonal axes from some reference point. In this coordinate system one can directly read physical intervals and distances from the coordinates and everything takes its familiar form. Moreover, it is expected that one would be allowed to define `observables' using this preferred coordinate system.

There are several objections that I want to raise against this line of reasoning. First, a clock hypothesis and a rod hypothesis are still in play, even if we are choosing a simple coordinate system. That is, we still need to postulate that $t$ correlates with the readings of clocks and that $x,y,z$ correlate with distance measurements along the $x,y,z$ directions. In this sense, the way coordinates acquire physical meaning is still by means of an assumption that links them with the behavior of reference objects, just as in general relativity. The meaning of the coordinates $t,x,y,z$ is therefore just the same as the meaning of an arbitrary coordinate system $x^{\mu}$ once the tensors $t_{\mu}$ and $h_{\mu\nu}$ are specified.

Second, the fact that the coordinate system is particularly simple and convenient implies that it is easy to give a physical interpretation to it, but for other coordinate systems there will also be some physical interpretation available, even if more complicated. For instance, we could build a coordinate system that corresponds to the times that a family of clocks, in some complicated state of motion, measure and with the times that some signals they exchange in between them take to travel. This coordinate system is not as straightforward to interpret as the standard one, but seems to be just as physical or meaningful, as we are still able to connect coordinates with measurements of some set of bodies. In other words, all coordinate systems get physical meaning by means of the clock hypothesis, and the fact that some may be more simple or convenient doesn't seem to make them physically privileged.

Third, in general relativity, at least for certain spacetimes, one could also choose a preferred system of coordinates by appealing to some simplicity or to their easy interpretability. Above I have commented on how some coordinates in Minkowski spacetime or in Schwarzschild spacetimes get an easy and straightforward interpretation. This would seem to imply that at least for some simple or symmetric spacetimes in general relativity, we can find physically meaningful spacetime coordinates\footnote{\citep{Marchetti2021} defend the position that for symmetric spacetimes one is able to make sense of general relativity by exploiting these symmetries, but that for generic spacetimes one has to deal with deep conceptual difficulties. Here I am arguing that for any spacetime, symmetric or not, we are able to give physical meaning to our coordinates.} in the same way we found them for Newtonian spacetimes. However, following what I have argued above, the claim that we can find physical meaning to coordinates in general relativity is extendable for arbitrary spacetimes, as the clock hypothesis will allow us to connect them, probably in a complicated manner, with the readings of some set of clocks, with signals sent from them, or some similar, presumably not trivial, and perhaps only locally valid construction.

Fourth, a different kind of argument for the position that there is something special about the case of Newtonian spacetime is that the set of coordinates $t,x,y,z$ is unique in that it is the only one (up to a temporal and spatial translation and a rotation) which `matches' the structures of spacetime. It is true that this coordinate system is unique in that the time coordinate is directly Newtonian absolute time and that the dynamical equations of motion take the familiar form of Newton's second law ($\vec{F}=m\frac{d^2\vec{x}}{dt^2}$). For general relativistic spacetimes there is no absolute time, no privileged foliation, and coordinates systems in which force laws take the simple form $\vec{F}=m\frac{d^2\vec{x}}{dt^2}$ can be defined only locally in general. However, this difference is a difference about the structures of spacetime and not about the meaning of the coordinates. The `privileged' Newtonian coordinates $t,x,y,z$, describe the same spacetime and contain the same information as any other coordinate system $x^{\mu}$ and the tensors $t_{\mu}$ and $h_{\mu\nu}$. This spacetime is different from a general relativistic spacetime, but the role of coordinates remains analogous.   

To complete the analogy let me comment that the GPS observables construction could also be implemented in the Newtonian spacetime case. Similarly to the case of general relativity, this construction allows defining a system of coordinates in relation to a system of satellites sending signals in spacetime, and authors like Rovelli would consider this coordinate system to be a legitimate and physical coordinate system.

Figure \ref{figure_gps_coord} which represented the idea for general relativity works also fine for representing how it could be implemented in a Newtonian spacetime. Just as in the case of general relativity, we would have four satellites traveling from a spacetime point $P$ at uniform speeds in four directions of space, each of them carrying their own clock and each of them sending signals with the reading of their clock at the time of emitting the signal. At every point in a region of spacetime the four signals $s^{\alpha}$ are received and they can be used for identifying the spacetime point. Taking into account the velocities of the satellites, that now the clocks measure absolute time, and that light signals now travel just at velocity $c$ with respect to absolute space, one is able to infer the relation between the coordinates $s^\alpha$ and the standard coordinates $t,x,y,z$. In this sense, we are able to express the geometry of Newtonian spacetime in terms of this system of `physical' coordinates.

Now one could ask, is this system of coordinates more `physical' than the $t,x,y,z$ coordinate system? Is it so in the case that there is no actual system of clocks of rods measuring $t,x,y,z$? Would it be so if the satellites hadn't been there? Is it more physical than an arbitrary system? As I have been arguing above, I take it that all coordinate systems are equally physical or meaningful, and that all get their physical meaning by means of the clock hypothesis that connects them with what clocks would measure, even if no such system existed in the world. The GPS coordinate system is a clever coordinate system with a straightforward implementation, but even without it, we are able to make sense of our spacetime theories, both Newtonian and general relativistic.

The analogy between Newtonian and general relativistic models shows that the claim by Rovelli that there is a difference between general relativity and Newtonian models in that reference objects need to be explicitly introduced in the case of general relativity and not in the case of Newtonian spacetime is mistaken. That is, in the same way we can treat clocks and rods externally and deduce their behavior from the coordinates $t,x,y,z$, from $s^{\alpha}, t_{\alpha},$ and $h_{\alpha\beta}$, or from $x^{\mu}, t_{\mu},$ and $h_{\mu\nu}$ in a Newtonian spacetime, we can do the same in the case of a general relativistic spacetime and treat clocks and reference objects externally. In the next section I will comment on the possible reactions from someone holding positions similar to Rovelli's, but for now let me take this to be a strong argument against claim 2.

In sections \ref{sect_intro_general_relativity} and \ref{sect_relationalism} I argued that there are substantivalist and relationalist interpretations of general relativity that allowed denying claim 1, i.e., the claim that we need to build physical observables in order to extract all the physical content of general relativity. In the Newtonian spacetime model case, the same two families of interpretations can be held and one can equally claim that there is no need to introduce physical observables or physical coordinate systems in order to have a sensible interpretation of such a model. The analogy shows that there is nothing in a general relativistic model which makes it the case that the interpretations of spacetime models that are valid in the Newtonian case are also valid in the general relativistic case. In this sense, if we have a complete interpretation of Newtonian models, we also have a complete interpretation of relativistic ones, and claim 1 is false. 

Similarly, the claim 3 that there are deep ontological differences between different spacetime models is clearly question-begging given the analogy between both types of models. We have seen how the different models have different causal, geometric, and inertial structures, but at this level there is no difference that would support the claims that one finds in the work of Rovelli, Rickles, and Earman cited above. 

All in all, the analysis of Newtonian spacetime models and their formulation in the language of differential geometry has made clear that the way we understand spacetime models is not different, or at least that it does not need to be, in the cases of general relativity and of Newtonian physics. This realization represents a challenge to the views I am opposing in this article and to the claims 1-3. In the next section I will analyze the ways my argument could be responded to.

\section{Possible replies by the defenders of `observables'}\label{sect_replies}

The previous sections have illustrated how at a kinematical level I find that the diffeomorphism invariance of differential geometrical models does not represent an obstacle at the time of giving a physical interpretation of such a model. In particular, some sort of clock hypothesis allows us to make a connection between coordinates and geometrical objects and what clocks and rods would measure. For this, there is no need to introduce any preferred or `physical' coordinate system or to define `observables'. In this section I will analyze two possible responses that a defender of `observables' may take in order to save part of the claims 1-3.

As I mentioned in section \ref{sect_need_for_obs}, the two motivations of the defenders of the need for `observables' pull in different directions when confronted with the analogy between the different types of spacetime models. While the arguments from the gauge analysis of general relativity can still be seen as pulling for the claim that there is a fundamental distinction between general relativistic and Newtonian spacetimes, the arguments from the radical relationalist perspective invite us to drop the claim that there is a difference. I will analyze both possible positions in turn.

\subsection{Does gauge symmetry make a difference?}\label{sect_gauge_difference}

The defender of the need for physical observables who motivates their position on their analysis of the diffeomorphism invariance of general relativity as a gauge symmetry could complain that my discussion in sections \ref{sect_GPS_observ_GR} and \ref{sect_Newtonian_differential_geometry} does not address their arguments. Moreover, they could accept the analogy proposed in the previous section, and even acknowledge that if we left aside the gauge aspect, the interpretation of general relativistic models and Newtonian models could be the same. However, from their point of view, analyzing general relativity as a gauge theory completely changes the picture and forces one to adopt a different interpretation. For instance, Earman claims that:
\begin{quote}
Physicists commonly take the substantive requirement of general covariance to mean that the laws exhibit diffeomorphism invariance \textit{and} that this invariance is a gauge symmetry. This latter requirement does place restrictions on the content of a spacetime theory. 

\citep[p. 1, his emphasis]{Earman2006}
\end{quote}
Earman (as well as Rickles and Rovelli in certain passages) introduces a distinction between theories that are expressed in the language of differential geometry and that therefore show some sort of diffeomorphism invariance and theories that show this invariance and for which this invariance is considered a gauge symmetry. In this sense, he distinguishes between formal general covariance (FGC) and substantive general covariance (SGC). Earman thinks there is a difference between a theory like Newtonian mechanics expressed in the language of differential geometry and general relativity, as the former would show FGC and the latter SGC. Given this supposed difference, Earman concludes that:
\begin{quote}
(C1) Since typical pre-general relativistic theories satisfy FGC, but not SGC, the general covariance of these theories does not rule out naive realism that takes the theory at face value as characterizing a world in terms of a manifold on which live various geometric object fields. (C2) For GTR and other spacetime theories that satisfy SGC, there are two immediate negative implications: (i) the so-called metrical essentialism is ruled out from the start since it is incompatible with diffeomorphism invariance as a gauge symmetry. (ii) Naive realism is also ruled out. 

\citep[p. 13]{Earman2006}
\end{quote}
As we see, the difference between FGC and SGC is supposed to be such that the interpretation of the models that have them would be radically different, despite all the apparent similarities. If that is right, then my argument in section \ref{sect_Newtonian_differential_geometry} would be blocked.

The key ingredient in Earman's argument is to claim that when the diffeomorphism invariance is considered a gauge symmetry this has important consequences for its interpretation. To be considered a gauge symmetry, according to Earman, it is not enough for it to be a transformation that maps a kinematical possible model to another one, as this would imply that models with an FGC would be misclassified as models with a gauge symmetry. Earman introduces the requirement that for a symmetry transformation to be considered a gauge transformation it needs to be a transformation that affects dynamical fields and variables and not fixed ones\footnote{Note that the sense of `dynamical' that these authors have in mind here is different from another possible sense, which is to define dynamical as changing in (space)time. The sense in which Earman is using dynamical is of interacting or dependent on other degrees of freedom. That is, spacetime is dynamical in this sense when it is true that if the matter distribution had been different, the properties of spacetime would have been different. This sense of `dynamical' implies the possibility of temporally changing spacetimes, but notice that the two senses are independent. That is, we can have spacetimes that are dynamical in both senses, only in one or in none.}. According to this definition, diffeomorphisms are gauge transformations in general relativity as they are symmetry transformations of dynamical variables ($g_{\mu\nu}, \phi$), but not in Newtonian mechanics, as the transformation affects dynamical variables ($\phi$), but also fixed variables ($h_{\mu\nu}, t_{\mu}$)\footnote{There are some subtleties regarding whether one can express theories like Newtonian mechanics in terms of action principles showing invariance under local diffeomorphisms, but I will leave them aside for the sake of the argument and refer the reader to \citep{pooley_background_2017} for a complete discussion of these details.}. If in the Newtonian case one works in a coordinate system like $t,x,y,z$ the possibility of defining a symmetry transformation mapping from this formulation to a formulation in terms of $h_{\mu\nu}$ and $t_{\mu}$ is not even explicit. 

Earman's definition is made in terms of local symmetries of an action principle, requiring that these local symmetries are transformations involving only dynamical variables and not fixed structures. Action principles with these symmetries have singular Lagrangians, and need to be treated as constrained systems in the Hamiltonian formalism. According to Earman's analysis, symmetries with these properties need to be interpreted in radically different ways from the way they are treated when the symmetries affect fixed structures. 

Does this difference between the formalization of the dynamics of different spacetime theories mean that there is some important difference at the time of interpreting their models? In my opinion, it does not. As I have just discussed, the reason why there is a formal difference between these models is just that while some models take spatiotemporal structures to be dynamical, others take them to be fixed. That is, the fact that in Newtonian mechanics one always has the same spacetime, while in general relativity one can have different models with different spacetimes is ultimately responsible for the different interpretations that authors like Earman propose for the two kinds of models. In the literature the term `background independence' is used to refer to the fact that spacetime is dynamical in general relativity and that there is no fixed structure in the theory. While I agree that this is a fact of the theory that makes it special and distinguishes it from other theories, I disagree with the conclusion that spacetime structures should be interpreted differently just because they are dynamical.

An argument supporting my claim is that for other physical variables we wouldn't change our interpretation of them depending on whether we consider them to be fixed or dynamical. Imagine a theory that describes the possible trajectories of a charged particle in an external electromagnetic field. Kinematical models of this theory are pairs $\langle A^{\mu}(x), x^{\mu} \rangle$. These models have what one intuitively would call a gauge symmetry, given that, as is well-known, different $A^{\mu}$s can represent the same electromagnetic field. Now, according to Earman's notion of gauge we need to distinguish between two situations. First, we can consider the electromagnetic field to be fixed as it is an external electromagnetic field that we can take to be given. In this sense, even if $A^{\mu}$ would appear in our models and in the action, it wouldn't be a dynamical field and the transformation that relates kinematically equivalent models would not count as a gauge transformation. Alternatively, we can consider the electromagnetic field to be a dynamical field and include in the action principle the terms that would give rise to Maxwell equations. In this case, according to Earman the transformation would count as a gauge transformation and this would have an impact on the way we interpret the theory. From the technical point of view, it is true that the electromagnetic field would satisfy the definition of `observable' as a phase space function which has vanishing Poisson brackets with the gauge generators only in the case that we treat it as a dynamical field. But do we want to claim that it is observable in the intuitive sense only in this case? Do we really want to further claim that its nature and the way we want to interpret it changes depending on whether we take it to be fixed or not? My position is that there is really no good reason for doing so and that we should in both cases interpret the field as a field that deviates charged particles from inertial movement. Similarly, I believe that spacetime models should be given the same interpretation independently of whether one takes the spacetime structure to be fixed or whether one allows for it to change from model to model. That is, dynamical or not, spacetime structure defines a causal and inertial structure and a geometry we can measure with physical objects.

In this sense, I find that whether a field is fixed or not does not affect the way we ought to interpret it. The only difference is of course that a dynamical field `reacts' or can react to the influences of other fields, and this is one of the key insights of general relativity. But then, if fixity or dynamicity shouldn't affect the way we interpret a structure, do we really want to claim that a formal property such as having a singular Lagrangian (or being formulated as a constrained Hamiltonian theory) can have as substantive implications as Earman and Rickles argue? In my opinion, this is an unattractive position.

As I mentioned in section \ref{sect_gauge_observables}, there are important formal and conceptual differences between the gauge transformations of electromagnetism and the diffeomorphism invariance of general relativity which make it the case that applying blindly recipes and definitions that make sense in the former case to the latter is completely question-begging. I refer the reader again to my discussion in that section and to the references therein for further arguments against analysis like Earman's. In this sense, I believe that one can very reasonably resist the arguments from the gauge analysis and argue that claims 1-3 are false.

\subsection{Accepting the analogy}

Alternatively, a defender of physical observables can take a perspective more inspired by the radical relationalist motivation and accept my claim that the formal analogy between spacetime models should also be accompanied by an analogy in the way they are to be interpreted. However, the radical relationalist could turn my argument around and use it to argue that physical observables are necessary not only in general relativity but also in Newtonian physics and in other spacetime theories. 

There are some passages in which Rovelli seems to go in this direction:
\begin{quote}
For Newton, the coordinates $\vec{x}$ that enter his main equation [Newton's second law] are the coordinates of absolute space. However, since we cannot directly observe space, the only way we can coordinatize space points is by using physical objects. [...] coordinates $\vec{x}$ [...] are therefore defined as distances from a chosen system $O$ of objects.[...]

In other words, the physical content of [Newton's second law]  is actually quite subtle: There exist reference objects $O$ with respect to which the motion of any other object $A$ is correctly described by [Newton's second law].

This is a statement that begins to be meaningful only when a sufficiently large number
of moving objects is involved. Notice also that for this construction to work it is important that the objects $O$ forming the reference frame are not affected by the motion of the object $A$. There
shouldn't be any dynamical interaction between $A$ and $O$.

\citep[pp. 87-88]{Rovelli2004}
\end{quote}

In this passage we see how Rovelli holds a relationalist position also for the case of Newtonian physics. As a relationalist, he rejects defining distances with respect to absolute space and he defines only distances between physical objects and a set of privileged reference objects which makes the equations of motion look simple. What makes Rovelli's position in this passage singular is that it requires the reference objects to exist, to follow inertial trajectories, and not to interact with the rest of the bodies. In this sense, we see that Rovelli claims that Newtonian coordinates are physical because they encode the distances with respect to this privileged set of physical objects. This is different not only from what a substantivalist would claim, but also from what many relationalists would claim. Standard relationalism does not need to postulate that there exist reference objects that are inertial and non-interacting, and it is able to accommodate the predictions of Newtonian physics for a set of $n$ bodies. That is, relationalism can hold that in order to predict the evolution of the distances between the $n$ bodies one can embed them in a Newtonian space-time, apply Newton's equations, extract the predictions, and forget about spacetime. In this sense, for this kind of relationalist, the position in space is only a tool that encodes the distances with respect to other bodies and allows computing their evolution, but does not represent a position in absolute space or the distance with respect to some privileged, existing body. The relationalist can interpret coordinates in a counterfactual way: if there existed privileged reference objects, coordinates $x,y,z$ would represent the distances of the rest of the bodies with respect to them.

We see that there is a parallelism again with the case of general relativity: I argued that spacetime models encode the distances with respect to reference objects (the satellites) even if they weren't explicitly there in the model. That is, according to plausible substantivalist and relationalist positions one could claim that coordinates encode distances with reference objects, but in both cases these coordinates can be understood in a counterfactual fashion. Meanwhile, Rovelli claims that in order for them to be physical, the reference objects need to be there in both situations. As I have argued, this is a possible position to hold, but not a very attractive one from my point of view.

We can also notice that in this passage Rovelli is implying that even if we want to claim that coordinates represent distances with respect to reference objects, one does not need to include these reference objects in the Newtonian model. That is, one takes the Newtonian model written in terms of the coordinate system $x,y,z$ to imply that there exist reference objects with respect to which the distances are $x,y,z$. But in the case of general relativity we saw how Rovelli rejected interpreting the coordinates $s^{\alpha}$ as encoding the relations with the satellites unless one explicitly included the satellites in the model. According to Rovelli, it was only when one included the satellites that one was able to extract all the predictions of the general relativistic model. To me, this difference in the way of treating the different spacetime models is not justified.

In the above quote by Rovelli we see that he demands that reference objects do not interact with the rest of the bodies we are studying. This seems to be a difference with the way he treats reference objects in general relativity, as is clear from the GPS example. But one can question this different way of treating reference objects in both cases. Having reference objects in Newtonian spacetime that do not interact at all with the rest of the bodies is clearly an idealization, as all the bodies in Newtonian mechanics interact gravitationally. In this sense, we never have an ideal set of bodies such that we can define coordinates with respect to them in a way that exactly matches Newtonian coordinates $t,x,y,z$. If we admit a degree of idealization regarding reference objects and their effect on other bodies in Newtonian physics, why not accept the same in the case of general relativity? In the example of the satellites it is true that the satellites and signals will interact gravitationally with the rest of matter and with the geometry of spacetime, but for light satellites and far away from them this effect will be small. It is therefore a very reasonable idealization to exclude their effect from our model, and in any case it is just analogous to the idealization made in Newtonian physics.

As I have been arguing in this article, I take it that this kind of analogy urges us to reject Rovelli's view about spacetime theories. However, Rovelli could take the tension just mentioned and take it to imply that his description of Newtonian spacetime in terms of a relationalism with respect to some idealized reference objects is inferior to an account in which no such idealized object is introduced. Indeed, in some other discussions of Newtonian mechanics, he seems to be arguing precisely for this claim:
\begin{quote}
According to Newton, we never directly measure the \textit{true} time variable $t$. Rather, we always construct devises, the ``clocks'' indeed, that have observable quantities (say, the angle $\beta$ between the clock’s hand and the direction of the digit ``12''), that move proportionally to the true time, within an approximation good enough for our purposes. In other words, we can say, following Newton, that what we can observe are the system’s quantities $a_i$ and the clock’s quantity $\beta$, and their relative evolution, namely the functions $a_i(\beta)$; but we describe this in our theory by assuming the existence of a ``true'' time variable $t$. We can then write evolution equations $a_i (t)$ and $\beta (t)$, and compare these with the observed change of $a_i$ with the clock’s hand $a_i (\beta)$.

Thus, it is true \textit{also} in non-relativistic mechanics that what we measure is only relative evolution between variables. But it turns out to be convenient to \textit{assume}, with Newton, that there exist a background variable $t$, such that all observables quantities evolve with respect to it, and equations are simple when written with respect to it. 

What I propose to do in the following is simply to drop this assumption.

\citep[p. 1479, his emphasis]{Rovelli2011}
\end{quote}
In this passage it is clear that Rovelli intends to apply to Newtonian spacetime the same interpretation that he makes of general relativity. In this sense, from Rovelli's perspective what we observe are relations between physical quantities, and hence one should avoid making reference to coordinate systems, time variables, or similar constructions. However, adopting such an interpretation does not make the stronger claims I have been discussing in this article true.

Let me start with claim 2. The quotation above undermines the claim that we need to explicitly introduce internal reference systems. Rovelli claims that it is correlations between the clock and the other system in his model that we observe, but then he discusses how one can introduce a variable $t$ that is convenient and makes equations simple. In this sense, what Rovelli is implying is that one can use the evolution of physical variables $a_i$ with respect to $t$ to predict what for him is truly physical and observable, $a_i(\beta)$. However, this allows treating clocks externally and it is precisely the way we usually treat them in Newtonian physics. In the case of general relativity the same applies, as the predictions like $\varphi(s^{\alpha})$ in the example of the satellites are written in terms of coordinates, but we are able to connect them with `physical correlations', even if we didn't include the clocks in our model.

Similarly, I find that claim 1 is not supported by this relationalist analysis. As Rovelli admits that the variable $t$ can encode the behavior of clocks in Newtonian spacetime, it seems that there is no obstacle to making a similar interpretation in general relativity. In this sense, even if one accepted Rovelli's claim that it is only correlations that are physical, one can reject the claim that we need to introduce them in our models in order to be able to extract the physical content of a general relativistic model. Moreover, the claim that we are far from capturing all the gauge invariant content of a general relativistic model is still question-begging.

Finally, even if Rovelli could formulate a consistent relationalist position, it seems clear to me that the claim that we are forced to adopt this interpretation in general relativity or in Newtonian physics is false. As I have been arguing in this article, there are attractive and consistent positions one can hold about space and time, for any spacetime theory, which give a perfectly fine and complete interpretation of spacetime theories. In this sense, positions like Rovelli's or Earman's are not forced into us as the only way of understanding spacetime theories, they are just one particular option, which I find is not as compelling as others.

\section{Conclusions}\label{sect_conclusions}

In this article I have argued against the view that we need `physical coordinate systems' or `observables' in order to have a complete interpretation of general relativity or to extract the physical content of a general relativistic model. I have argued that there is an analogy between general relativistic spacetimes and other spacetimes and that this should make us give the same interpretation to the different spacetimes appearing in our physical theories. That is, spacetime defines a causal structure, a geometric structure that we measure with clocks and rods, and an inertial structure that describes how free bodies move. In different theories we find different spatiotemporal structures, but these three basic interpretative tenets hold for the different spacetimes. In this sense, I have argued that we have perfectly fine interpretations of spacetime models, including general relativity, with no need to introduce this notion of physical observable.

I have also analyzed the two motivations for adopting the `physical observables' view, and I have found that there are serious worries associated with them that make the `physical observables' view not attractive and certainly not the only way of interpreting spacetime models. First, I have found that the gauge analysis of general relativity is based on an analogy with gauge theories like electromagnetism which can be challenged given the disanalogies between the two types of transformation. Moreover, I have argued that this view seems to be too formalistic in that it follows blindly some formal recipes to deduce what for me are quite absurd conclusions such as that a physical structure should receive a different interpretation depending on whether we take it to be fixed or dynamical. Second, the kind of relationalism defended by Rovelli can be questioned and is certainly not forced into us by the structure of general relativity or other spacetime theories, but even from that perspective I have found that the claim that we do not have a complete interpretation of spacetime models if we do not explicitly introduce reference objects is false. 

\section*{Acknowledgments}

I want to thank an anonymous reviewer for their detailed and constructive comments that have been of invaluable help in improving this article.

\printbibliography

@article{Gryb2016,
	title = {Time remains},
	volume = {67},
	issn = {14643537},
	doi = {10.1093/bjps/axv009},
	number = {3},
	journal = {British Journal for the Philosophy of Science},
	author = {Gryb, Sean and Thébault, Karim P.Y.},
	year = {2016},
	pages = {663--705},
}

@incollection{Thebault2021,
	title = {The {Problem} of {Time}},
	url = {https://www.routledge.com/The-Routledge-Companion-to-Philosophy-of-Physics/Knox-Wilson/p/book/9781138653078#},
	urldate = {2021-07-12},
	booktitle = {The {Routledge} {Companion} to {Philosophy} of {Physics}},
	publisher = {Routledge},
	author = {Thébault, Karim P.Y.},
	editor = {Knox, Eleanor and Wilson, Alastair},
	month = sep,
	year = {2021},
}

@article{Gryb2010,
	title = {Jacobi’s principle and the disappearance of time},
	volume = {81},
	url = {https://journals.aps.org/prd/abstract/10.1103/PhysRevD.81.044035},
	doi = {10.1103/PhysRevD.81.044035},
	number = {4},
	urldate = {2021-10-14},
	journal = {Physical Review D},
	author = {Gryb, Sean},
	month = feb,
	year = {2010},
	note = {Publisher: American Physical Society},
	pages = {044035},
}

@book{Brown2006,
	title = {Physical {Relativity}: {Space}-time structure from a dynamical perspective},
	isbn = {978-0-19-160391-4},
	urldate = {2021-05-13},
	publisher = {Oxford University Press},
	author = {Brown, Harvey R.},
	month = sep,
	year = {2006},
	doi = {10.1093/0199275831.001.0001},
	note = {Publication Title: Physical Relativity: Space-time structure from a dynamical perspective},
}

@article{Marchetti2021,
	title = {Effective relational cosmological dynamics from quantum gravity},
	volume = {2021},
	issn = {10298479},
	url = {https://doi.org/10.1007/JHEP05},
	doi = {10.1007/JHEP05(2021)025},
	number = {5},
	urldate = {2021-06-21},
	journal = {Journal of High Energy Physics},
	author = {Marchetti, Luca and Oriti, Daniele},
	month = may,
	year = {2021},
	note = {arXiv: 2008.02774
Publisher: Springer Science and Business Media Deutschland GmbH},
	pages = {25},
}

@article{Hoefer1996,
	title = {The {Metaphysics} of {Space}-{Time} {Substantivalism}},
	volume = {93},
	url = {https://www.pdcnet.org/pdc/bvdb.nsf/purchase?openform&fp=jphil&id=jphil_1996_0093_0001_0005_0027},
	doi = {10.2307/2941016},
	number = {1},
	urldate = {2022-04-01},
	journal = {The Journal of Philosophy},
	author = {Hoefer, Carl},
	month = jan,
	year = {1996},
	note = {Publisher: Philosophy Documentation Center},
	pages = {5--27},
}

@misc{Norton2019,
	title = {The {Hole} {Argument}},
	url = {https://plato.stanford.edu/archives/sum2019/entries/spacetime-holearg/},
	journal = {The \{Stanford\} Encyclopedia of Philosophy},
	author = {Norton, John D and Zalta, Edward N},
	year = {2019},
	note = {Publisher: Metaphysics Research Lab, Stanford University
Medium: {\textbackslash}url\{https://plato.stanford.edu/archives/sum2019/entries/spacetime-holearg/\}},
}

@article{Earman2002,
	title = {Thoroughly {Modern} {Mctaggart}: {Or}, {What} {Mctaggart} {Would} {Have} {Said} {If} {He} {Had} {Read} the {General} {Theory} of {Relativity}},
	volume = {2},
	number = {3},
	journal = {Philosophers' Imprint},
	author = {Earman, John},
	year = {2002},
	pages = {1--28},
}

@article{Earman2006,
	title = {The {Implications} of {General} {Covariance} for the {Ontology} and {Ideology} of {Spacetime}},
	volume = {1},
	issn = {18711774},
	doi = {10.1016/S1871-1774(06)01001-1},
	number = {C},
	journal = {Philosophy and Foundations of Physics},
	author = {Earman, John},
	year = {2006},
	pages = {3--23},
}

@article{Rovelli2011,
	title = {"{Forget} time" {Essay} written for the {FQXi} contest on the {Nature} of {Time}},
	volume = {41},
	issn = {00159018},
	url = {https://link-springer-com.are.uab.cat/article/10.1007/s10701-011-9561-4},
	doi = {10.1007/s10701-011-9561-4},
	number = {9},
	urldate = {2021-01-19},
	journal = {Foundations of Physics},
	author = {Rovelli, Carlo},
	month = sep,
	year = {2011},
	note = {Publisher: Springer},
	pages = {1475--1490},
}

@article{Pitts2017,
	title = {Equivalent theories redefine {Hamiltonian} observables to exhibit change in general relativity},
	volume = {34},
	issn = {13616382},
	doi = {10.1088/1361-6382/aa5ce8},
	number = {5},
	journal = {Classical and Quantum Gravity},
	author = {Pitts, J. Brian},
	year = {2017},
	pages = {1--23},
}

@book{Rovelli2004,
	address = {Cambridge},
	title = {Quantum {Gravity}},
	isbn = {978-0-521-71596-6},
	url = {https://www.cambridge.org/core/product/identifier/9780511755804/type/book},
	urldate = {2019-12-10},
	publisher = {Cambridge University Press},
	author = {Rovelli, Carlo},
	month = nov,
	year = {2004},
	doi = {10.1017/CBO9780511755804},
}

@article{Maudlin2002,
	title = {Thoroughly {Muddled} {Mctaggart}: {Or}, {How} to {Abuse} {Gauge} {Freedom} to {Create} {Metaphysical} {Monstrosities}},
	volume = {2},
	number = {4},
	journal = {Philosophers' Imprint},
	author = {Maudlin, Tim W.E.},
	year = {2002},
	pages = {1--23},
}

@article{Pons2010,
	title = {Observables in classical canonical gravity: {Folklore} demystified},
	volume = {222},
	issn = {1742-6596},
	url = {https://iopscience.iop.org/article/10.1088/1742-6596/222/1/012018},
	doi = {10.1088/1742-6596/222/1/012018},
	number = {1},
	journal = {Journal of Physics: Conference Series},
	author = {Pons, Josep M. and Salisbury, D. C. and Sundermeyer, Kurt A.},
	month = apr,
	year = {2010},
	pages = {012018},
}

@article{Pitts2018,
	title = {Equivalent {Theories} and {Changing} {Hamiltonian} {Observables} in {General} {Relativity}},
	volume = {48},
	issn = {15729516},
	url = {https://doi.org/10.1007/s10701-018-0148-1},
	doi = {10.1007/s10701-018-0148-1},
	number = {5},
	journal = {Foundations of Physics},
	author = {Pitts, J. Brian},
	year = {2018},
	note = {Publisher: Springer US},
	pages = {579--590},
}

@article{MozotaFrauca2023,
	title = {Reassessing the problem of time of quantum gravity},
	volume = {55},
	issn = {0001-7701},
	url = {https://arxiv.org/abs/2301.07973v1},
	doi = {10.1007/s10714-023-03067-x},
	number = {1},
	urldate = {2023-01-20},
	journal = {General Relativity and Gravitation},
	author = {Mozota Frauca, Álvaro},
	month = jan,
	year = {2023},
	note = {arXiv: 2301.07973},
	pages = {21},
}

@article{Thebault2012,
	title = {Three denials of time in the interpretation of canonical gravity},
	volume = {43},
	issn = {1355-2198},
	doi = {10.1016/J.SHPSB.2012.09.001},
	number = {4},
	urldate = {2023-01-24},
	journal = {Studies in History and Philosophy of Science Part B: Studies in History and Philosophy of Modern Physics},
	author = {Thébault, Karim P.Y.},
	month = nov,
	year = {2012},
	note = {Publisher: Pergamon},
	pages = {277--294},
}

@incollection{Rovelli2022,
	title = {Philosophical {Foundations} of {Loop} {Quantum} {Gravity}},
	isbn = {2211.06718v2},
	url = {https://arxiv.org/abs/2211.06718v2},
	urldate = {2023-03-20},
	author = {Rovelli, Carlo and Vidotto, Francesca},
	month = nov,
	year = {2022},
	note = {arXiv: 2211.06718},
}

@article{Rovelli1991,
	title = {What is observable in classical and quantum gravity?},
	volume = {8},
	issn = {0264-9381},
	url = {https://iopscience.iop.org/article/10.1088/0264-9381/8/2/011},
	doi = {10.1088/0264-9381/8/2/011},
	number = {2},
	urldate = {2023-04-21},
	journal = {Classical and Quantum Gravity},
	author = {Rovelli, Carlo},
	month = feb,
	year = {1991},
	note = {Publisher: IOP Publishing},
	pages = {297},
}

@article{Rovelli2002a,
	title = {{GPS} observables in general relativity},
	volume = {65},
	issn = {05562821},
	url = {https://journals-aps-org.are.uab.cat/prd/abstract/10.1103/PhysRevD.65.044017},
	doi = {10.1103/PhysRevD.65.044017},
	number = {4},
	urldate = {2023-05-04},
	journal = {Physical Review D},
	author = {Rovelli, Carlo},
	month = jan,
	year = {2002},
	note = {arXiv: gr-qc/0110003
Publisher: American Physical Society},
	pages = {044017},
}

@article{Bergmann1961,
	title = {Observables in {General} {Relativity}},
	volume = {33},
	issn = {00346861},
	url = {https://journals-aps-org.are.uab.cat/rmp/abstract/10.1103/RevModPhys.33.510},
	doi = {10.1103/RevModPhys.33.510},
	number = {4},
	urldate = {2023-06-19},
	journal = {Reviews of Modern Physics},
	author = {Bergmann, Peter G.},
	month = oct,
	year = {1961},
	note = {Publisher: American Physical Society},
	pages = {510},
}

@article{Knox2014,
	title = {Newtonian {Spacetime} {Structure} in {Light} of the {Equivalence} {Principle}},
	volume = {65},
	issn = {14643537},
	url = {https://www-journals-uchicago-edu.are.uab.cat/doi/10.1093/bjps/axt037},
	doi = {10.1093/BJPS/AXT037},
	number = {4},
	urldate = {2023-06-22},
	journal = {https://doi-org.are.uab.cat/10.1093/bjps/axt037},
	author = {Knox, Eleanor},
	month = dec,
	year = {2014},
	note = {Publisher: The University of Chicago Press},
	pages = {863--880},
}

@article{Kuchar1980,
	title = {Gravitation, geometry, and nonrelativistic quantum theory},
	volume = {22},
	issn = {05562821},
	url = {https://journals-aps-org.are.uab.cat/prd/abstract/10.1103/PhysRevD.22.1285},
	doi = {10.1103/PhysRevD.22.1285},
	number = {6},
	urldate = {2023-06-22},
	journal = {Physical Review D},
	author = {Kuchař, Karel V.},
	month = sep,
	year = {1980},
	note = {Publisher: American Physical Society},
	pages = {1285},
}

@book{Malament2012,
	title = {Topics in the {Foundations} of {General} {Relativity} and {Newtonian} {Gravitation} {Theory}.},
	isbn = {978-0-226-50247-2},
	url = {https://press.uchicago.edu/ucp/books/book/chicago/T/bo12893557.html},
	urldate = {2023-06-22},
	publisher = {University of Chicago Press},
	author = {Malament, David B.},
	year = {2012},
}

@book{Maudlin2012,
	title = {Philosophy of {Physics}: {Space} and {Time}},
	volume = {5},
	isbn = {978-0-691-14309-5},
	url = {http://www.jstor.org/stable/j.ctvc77bdv},
	urldate = {2023-06-22},
	publisher = {Princeton University Press},
	author = {Maudlin, Tim W.E.},
	year = {2012},
}

@article{Meskhidze2023,
	title = {Torsion in the {Classical} {Spacetime} {Context}},
	url = {https://arxiv.org/abs/2304.11248v1},
	urldate = {2023-09-01},
	author = {Meskhidze, Helen and Weatherall, James Owen},
	month = apr,
	year = {2023},
	note = {arXiv: 2304.11248},
}

@article{Earman1987,
	title = {What {Price} {Spacetime} {Substantivalism}? {The} {Hole} {Story}},
	volume = {38},
	issn = {00070882},
	url = {https://www.journals.uchicago.edu/doi/10.1093/bjps/38.4.515},
	doi = {10.1093/BJPS/38.4.515},
	number = {4},
	urldate = {2023-09-01},
	journal = {https://doi.org/10.1093/bjps/38.4.515},
	author = {Earman, John and Norton, John},
	month = dec,
	year = {1987},
	note = {Publisher: The University of Chicago Press},
	pages = {515--525},
}

@article{fletcher_light_2013,
	title = {Light {Clocks} and the {Clock} {Hypothesis}},
	volume = {43},
	issn = {1572-9516},
	url = {https://doi.org/10.1007/s10701-013-9751-3},
	doi = {10.1007/s10701-013-9751-3},
	language = {en},
	number = {11},
	urldate = {2024-01-25},
	journal = {Foundations of Physics},
	author = {Fletcher, Samuel C.},
	month = nov,
	year = {2013},
	pages = {1369--1383},
}

@incollection{pooley_background_2017,
	address = {New York, NY},
	series = {Einstein {Studies}},
	title = {Background {Independence}, {Diffeomorphism} {Invariance} and the {Meaning} of {Coordinates}},
	isbn = {978-1-4939-3210-8},
	url = {https://doi.org/10.1007/978-1-4939-3210-8_4},
	language = {en},
	urldate = {2024-01-25},
	booktitle = {Towards a {Theory} of {Spacetime} {Theories}},
	publisher = {Springer},
	author = {Pooley, Oliver},
	editor = {Lehmkuhl, Dennis and Schiemann, Gregor and Scholz, Erhard},
	year = {2017},
	doi = {10.1007/978-1-4939-3210-8_4},
	pages = {105--143},
}

@incollection{rickles_chapter_2008,
	series = {The {Ontology} of {Spacetime} {II}},
	title = {Chapter 7 {Who}'s {Afraid} of {Background} {Independence}?},
	volume = {4},
	url = {https://www.sciencedirect.com/science/article/pii/S1871177408000077},
	urldate = {2024-01-25},
	booktitle = {Philosophy and {Foundations} of {Physics}},
	publisher = {Elsevier},
	author = {Rickles, Dean},
	editor = {Dieks, Dennis},
	month = jan,
	year = {2008},
	doi = {10.1016/S1871-1774(08)00007-7},
	pages = {133--152},
}

@incollection{read_classical_2023,
	title = {Classical {Theories} of {Spacetime}},
	isbn = {978-0-19-288911-9},
	url = {https://doi.org/10.1093/oso/9780192889119.003.0004},
	urldate = {2024-01-25},
	booktitle = {Background {Independence} in {Classical} and {Quantum} {Gravity}},
	publisher = {Oxford University Press},
	author = {Read, James},
	editor = {Read, James},
	month = nov,
	year = {2023},
	doi = {10.1093/oso/9780192889119.003.0004},
	pages = {0},
}

@article{pooley_hole_2006,
	title = {A hole revolution, or are we back where we started?},
	volume = {37},
	issn = {1355-2198},
	url = {https://www.sciencedirect.com/science/article/pii/S1355219806000116},
	doi = {10.1016/j.shpsb.2005.11.003},
	number = {2},
	urldate = {2024-02-09},
	journal = {Studies in History and Philosophy of Science Part B: Studies in History and Philosophy of Modern Physics},
	author = {Pooley, Oliver},
	month = jun,
	year = {2006},
	pages = {372--380},
}

@article{kretschmann_uber_1917,
	title = {Über den {Physikalischen} {Sinn} der {Relativitätspostulate}},
	volume = {53},
	journal = {Annalen Der Physik},
	author = {Kretschmann, E.},
	year = {1917},
	pages = {575--614},
}

\end{document}